\documentclass[9pt,twocolumn,twoside]{osajnl}

\journal{ol} 

\setboolean{shortarticle}{true} 

\title{Turbulence correction with artificial neural networks}

\author[1,+]{Sanjaya Lohani}
\author[1,*]{Ryan T. Glasser}

\affil[1]{Tulane University, New Orleans, LA 70118, USA}

\affil[+]{slohani@tulane.edu}
\affil[*]{rglasser@tulane.edu}

\ociscodes{(010.1330) Atmospheric turbulence; (010.1080) Active or adaptive optics; (100.4996) Pattern recognition, neural networks; (060.4510) Optical communications; (200.2605) Free-space optical communication; (200.4260) Neural networks.}

\begin{abstract}
We design an optical feedback network making use of machine learning techniques and demonstrate via simulations its ability to correct for the effects of turbulent propagation on optical modes.  This artificial neural network scheme only relies on measuring the intensity profile of the distorted modes, making the approach simple and robust. The network results in the generation of various mode profiles at the transmitter that, after propagation through turbulence, closely resemble the desired target mode.  The corrected optical mode profiles at the receiver are found to be nearly identical to the desired profiles, with near-zero mean square error indices. We are hopeful that the present results combining the fields of machine learning and optical communications will greatly enhance the robustness of free-space optical links.
\end{abstract}

\setboolean{displaycopyright}{false}

\begin{document}

\maketitle

Free space optical communications may be utilized to establish high-bit-rate communications channels\cite{malik2015free}. Features such as large bandwidths, license free spectra, established security protocols, and many more\cite{shaulov_simulation-assisted_2005,sahbudin_performance_2013,willebrand_fiber_2001,kumar_impact_2013}, have resulted in the wide application of such optical communications links in military and civilian protocols alike\cite{chan_free-space_2006}. Recently, orthogonal spatial states of light, in particular those containing nonzero orbital angular momentum, have given rise to increased system capacities and even larger bit-transfer rates per photon, by employing various multiplexing and de-multiplexing techniques\cite{milione_4x_2015,zhao_high-base_2015,wang_terabit_2012,gbur_vortex_2008}. Despite these benefits, these optical communication schemes typically experience loss and signal fading caused by absorption, scattering, and turbulent propagation\cite{singh_performance_2013,zabidi_investigating_2011,fadhil_optimization_2013}. Among these effects, distortion due to atmospheric turbulence leads to cross-talk between the channels that may seriously degrade the bit error rate, and may easily destroy any classical or quantum advantage of the system\cite{rodenburg_influence_2012,capraro2012impact}. One attempt to mitigate the effects of turbulence is to construct an adaptive optical network which, in general, requires a complex setup and additional optical components\cite{rukosuev_adaptive_2015,li_adaptive_2017,marron_atmospheric_2009,zhu_free-space_2002,gladysz2015adaptable}, and unlike the present network does not actively provide feedback to generate robust optical mode profiles at the transmitter, which we show results in nearly ideal modes at the receiver.

\begin{figure}[b!] 
\centering\includegraphics[width=.72\linewidth]{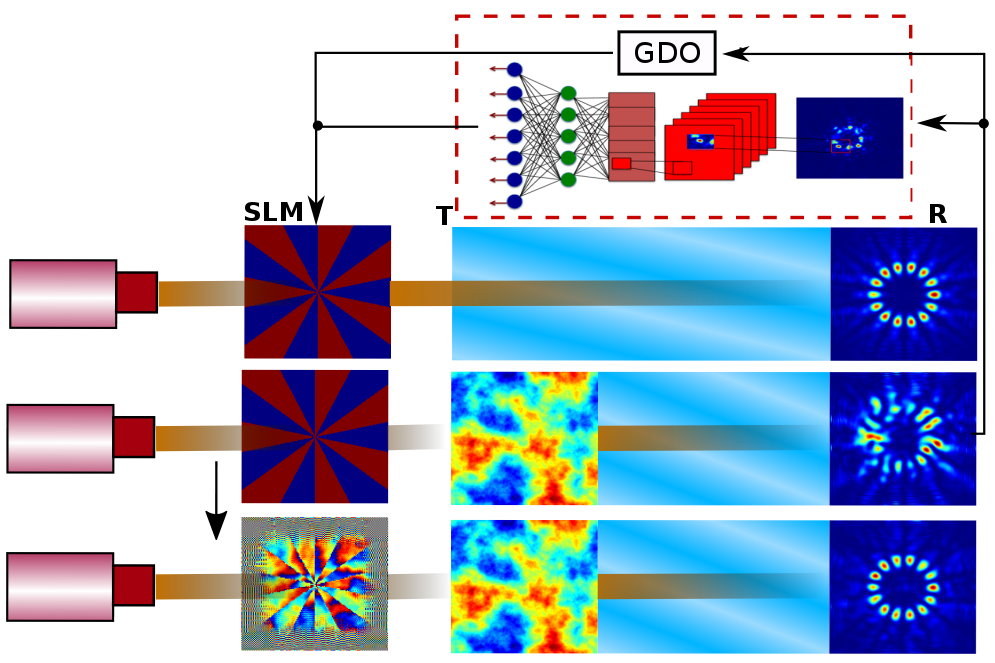}
\caption{Schematic of the turbulence mitigating network which is comprised of a spatial light modulator (SLM), transmitter (T), communication channel with turbulence, receiver (R), and a feedback network with a CNN and GDO.  The optical mode profiles on the right correspond to the desired image (top), distorted image due to turbulence (middle), and turbulence-corrected image at the receiver (bottom). }
\label{fig:Figure 1}
\end{figure}
In order to apply machine learning (ML) to correct for the simulated turbulent propagation of optical states, we have designed a customized convolutional neural network (CNN)\cite{MN_book} and a gradient descent optimizer (GDO)
, which serve as feedback and correction units. The simulation results presented here consist of multiple steps, which are visualized in Fig. \ref{fig:Figure 1}.
First, we simulate a laser directed onto a spatial-light modulator (SLM) with a given phase profile, $\Theta^{(\ell,-\,\ell)}$, which is used to convert each incident optical beam from a Gaussian mode profile into a Laguerre-Gauss mode of $\pm\, l$, where $l$ corresponds to the degree of orbital angular momentum (OAM) mode contains. The input Gaussian beam diameter and intensity are fixed throughout the simulations.  The resulting mode profile is then a “petal pattern” of $2l$ bright and dark spots, in a circular configuration. Note that we use the Fresnel Transfer Function (TF) propagator and Fresnel Impulse Response (IR) propagator\cite{voelz2011computational} to propagate the beam through the setups. We then use a Kolmogorov phase with the Von Karman spectrum effects\cite{bos_anisotropic_2015} model to simulate turbulence (whose strength is quantified by the standard refractive index, $C_n^2$)  in the atmosphere, which is given by Eq. \ref{eqn:phi_xy}
\begin{equation}
\Phi (X,Y) = \Re \{\mathcal{F}^{-1}(\mathbb{C}_{NN}\sqrt{\Phi_{NN}(\kappa)})\},
\label{eqn:phi_xy}
\end{equation}
with $\Phi_{NN}(\kappa)\,=\,0.023r_0^{-5/3}(\kappa^2 + \kappa_0^2)^{-11/6}\exp{(-\kappa^2/\kappa_m^2)} $, and a Fried parameter ($r_0$)\,=\,$(0.423k^2 C_n^2 Z)^{-3/5}$.  Here X-Y are Cartesian co-ordinates, $\Re$ corresponds to taking only the real part of the function, $\mathcal{F}^{-1}$ is the inverse fast Fourier transformation, $\mathbb{C}_{NN}$ are complex random normal numbers with zero mean and unit variance, sub-script ``NN'' represents the distribution over a sampling grid of size N\,$\times$\,N, with all other parameters given in Table \ref{tab: table1}. Note that we use a single turbulence phase screen to mimic the given atmospheric effects in the communication channel at a time. Thus, two different turbulence phase screens represent two different atmospheric turbulence channels (or turbulence). Next we simulate shining a Gaussian beam, $G(X,Y,w_0)$, through the channel and obtain a resultant intensity pattern at the receiver. 
For the unknown turbulence, $\Phi_{real}(X,Y)$, the pre-trained CNN, discussed in the next paragraph, makes a prediction of an unknown $C_n^2$ for a fedback, undesirable intensity pattern within 30 ms, and the corresponding turbulence pattern $\Phi_{est}(X,Y)$ is estimated. Due to the randomness of both $\mathbb{C}_{NN}$ and the test phase screen selection, as well as a limited number of training classes, the predicted $\Phi_{est}(X,Y)$ may not represent the exact unknown turbulence in the path. This requires the GDO, which optimizes $\mathbb{C}_{NN}$, and hence  $\Phi_{est}(X,Y)$, by minimizing the mean square error between the target and received intensity pattern within a few hundred iterations using a gradient descent algorithm. Note that we have derived a relation given by Eq. \ref{eqn:slm_phase} which is implemented in the feedback network to update the phase mask at the SLM.
\begin{equation}
\begin{split}
\Theta^j(X,Y) = \angle \Bigg[ \mathcal{F}^{-1}\bigg\{\frac{1}{H_1}\,\times\,\mathcal{F}\Big[\mathcal{F}^{-1}\big(\mathcal{F}(G(X,Y,w_0)\\
\exp(i\,\Theta^{(\ell,-\,\ell)}))\,\times\, H_{1}\big)\exp(-\,i\,\Phi_{est}^{j}(X,Y))\Big]\bigg\}\Bigg] 
\end{split}
\label{eqn:slm_phase}
\end{equation}
Here $\Theta^j(X,Y)$ is the updated phase mask at the $j^{th}$ iteration,  ``$\angle$'' is the arctan of the ratio of the imaginary part to the real part and $H_1$ is the transfer function between the SLM and transmitter plane. This results in a superposition pattern to be sent back through the turbulent medium. This new mode is then significantly more robust to propagation through the turbulent medium, resulting in a more ideal mode at the receiver, as shown in Fig. \ref{fig:Figure 1}.
\begin{table}[t!]
\noindent\makebox[\linewidth]{\rule{\linewidth}{0.6pt}}
\centering
\resizebox{\linewidth}{!}{
\begin{tabular}{ p{12em} p{4.15em}|p{15.54em}p{12em}}
Grid size (N$\times$N)&$128\times128$&Sample grid length (dX)&$4\times10^{-4}$\\
Wavelength ($\lambda$)&1550 nm&Gaussian beam waist ($\omega_0$)& 7 mm\\

Inner scale of turbulence ($l_{min}$)&1 mm&Outer scale of turbulence ($l_{max}$)&25 m\\

Wave-vector (k)&$2\pi/\lambda$&Spatial frequency ($\kappa$) at $l_{min}$ and $l_{max}$&$\kappa_{m}\,=\,5.92/l_{min}$,\quad$\kappa_0\,=\,2\pi/l_{max} $\\

Distance between SLM and transmitter & 1 m&Turbulence strength ($C_n^2$)&varies from $5\times10^{-12}m^{-2/3}$ to $9\times10^{-11}m^{-2/3}$\\
Distance between transmitter and receiver (Z)&25 m&Length of screens (L) & 51.2 mm ($>3\times 2\omega_0$)\\
\end{tabular}}
\caption{Parameters used to simulate propagation through the setups.}
\label{tab: table1}
\noindent\makebox[\linewidth]{\rule{\linewidth}{0.6pt}}
\end{table}

The network uses a convolution of a 5\,$\times$\,5 filter and 2\,$\times$\,2 max pooling, 100 fully connected neurons, and finally, a softmax layer at the output to locally build the 5 layer CNN as shown in Fig. \ref{fig:Figure 1}. This network is computationally efficient, even at local CPU stations, due to its small size and robustness, as investigated in\cite{deep_paper}. In order to train the CNN, we generate images with a resolution of 128\,$\times$\,128 using the Fresnel propagator (TF and IR) mentioned above for each value of $C_n^2$ given in Table \ref{tab: table1} 
with a common difference of $0.2179\,\times 10^{-11}$ m$^{-2/3}$, for a total of 40 different training classes. Because of $\mathbb{C}_{NN}$ in Eq. \ref{eqn:phi_xy}, each intensity image at the receiver is different from others even when they are from the same training class or have the same strength of turbulence. Note that we use the raw data of the images to train the CNN. Additionally the strength ($C_n^2$) of a test turbulence phase screen is random and uniformly chosen in the range of $0.5 - 9\times10^{-11}$m$^{-2/3}$, which is different from those in the training classes, for all the calculations except when finding the accuracy versus $C_n^2$ (in which case it is varied, in order to find the MSE as a function of $C_n^2$). Note that none of the training sets include the test patterns.


Prior to running the turbulence-correcting simulations, first we optimize the learning rate ($\eta$) of the GDO, number of optimizing iterations, and number of images per training class of the CNN, in order to increase network performance and computational efficiency. Additionally, we individually run the optimization 10 times for each set of parameters with respect to the accuracy at the receiver, in order to gather statistics.  Accuracy herein is defined as the measure of mean-squared error indices (MSE) between the received mode and the desired (target) mode intensity profiles. Note that we generate 600 intensity images at the receiver per training class, for a total of 24,000 images.  
\begin{figure}[b!] 
\centering\includegraphics[width=.9\linewidth]{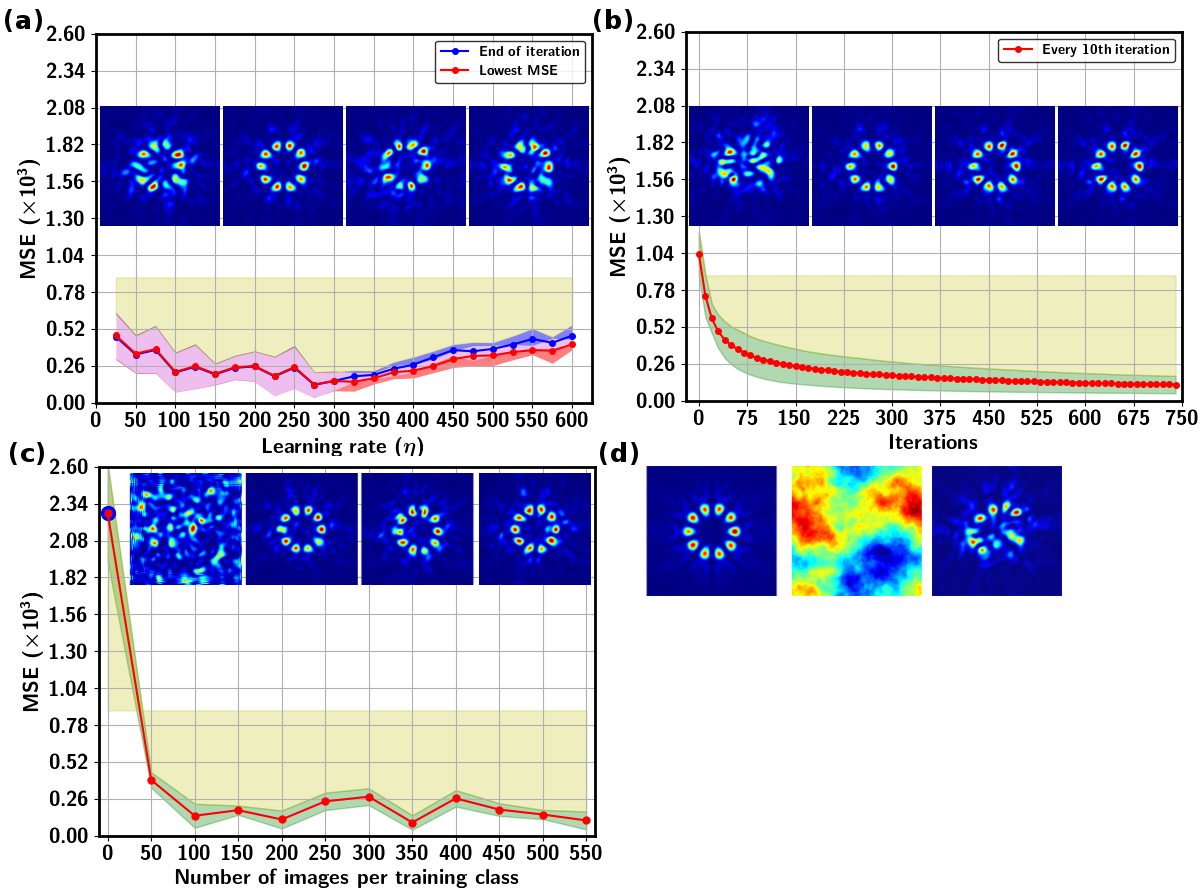}
\caption{
Optimizing the \textbf{(a)} learning rate of the GDO, \textbf{(b)} number of iterations, and \textbf{(c)} number of images per training class of the CNN. The translucent bands around the curves (red and blue in \textbf{(a)} and green in \textbf{(b,c)}) represent one standard deviation, whereas the magenta band is a common region between them. The yellow regions \textbf{(a,b,c)} represent the difference between the corrected and uncorrected MSE for the test image. \textbf{(d)} left: target superposition OAM for $\ell\,=\,\pm\,5$ intensity pattern, center: turbulence phase map, and right: distorted pattern at receiver. Examples of turbulence-corrected intensity patterns are shown in the insets from left to right, respectively, at \textbf{(a)} $\eta$: 25, 275, 450, and 600, \textbf{(b)} iterations: 0, 300, 500, 700, and \textbf{(c)} number of images per training class: 0, 200, 400, 500.}
\label{fig:Figure 2}
\end{figure}

In order to optimize $\eta$, we randomly select 200 images out of 600 images per class and train the CNN with them. After that the network setups are run up to 300 iterations with a pre-trained CNN to optimize $\eta$ as shown in Fig. \ref{fig:Figure 2} \textbf{(a)}. The blue and red curves represent the mean square error (MSE) for the final images after the end of 300th iteration and the lowest MSE among the entire 300 iterations, respectively. 
Note that we are only concerned with the lowest MSE (red curve) regime for the network optimization. We emphasize here that a lower MSE signifies a higher accuracy, or a more ideal image at the receiver. 
All of the MSE indices presented in this paper are of the order $10^3$. The images corresponding to the red curve in Fig. \ref{fig:Figure 2}\textbf{(a)} have a maximum MSE of 0.476 at $\eta\,=\,25$, and a minimum MSE of 0.131 at $\eta\,=\,275$, with standard deviations of 0.162 and 0.090, respectively. The MSE gradually decreases until $\eta$ reaches the optimal value of 275. Example images are shown in the insets, and Fig. \ref{fig:Figure 2} (d) shows the undistorted optical mode, turbulence phase map, and the uncorrected (distorted) mode. After taking into account the oscillations, standard deviations, and final accuracies, we set $\eta\,=\,275$ as the optimized learning rate. 

\begin{figure}[h!]
\centering\includegraphics[width=.92\linewidth]{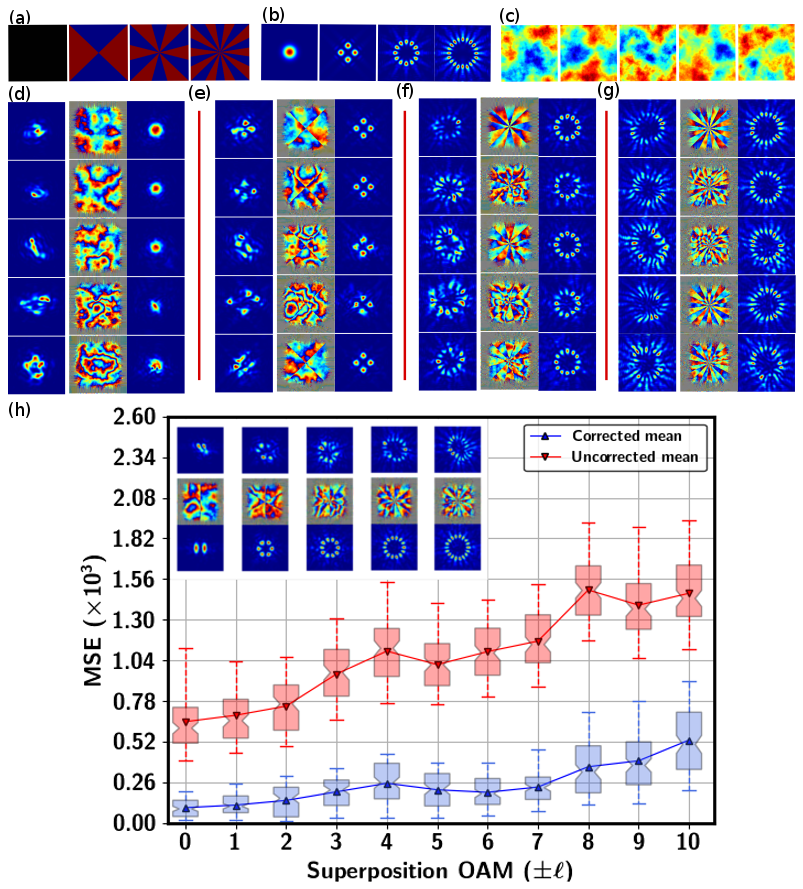}
\caption{Examples of \textbf{(a)} ideal phase masks on the SLM for $\pm \ell$ = 0, 2,6, and 10, \textbf{(b)} target intensity patterns at receiver (without any turbulence), \textbf{(c)} turbulence phase screens. Similarly, from left to right, respectively: \textbf{(d, e, f, g)} distorted intensity patterns in the presence of turbulence shown in ``(c)'', the corresponding updated phase mask at SLM that is found, and the corrected OAM image at receiver, after turbulent propagation. Finally, \textbf{(h)} MSE versus various superposition OAM modes, from $\pm \ell\,=\,0\,-\,10$, after propagation through turbulence of the same strength, but of randomly-varying phase distributions (100 different images per each $\ell$). The  blue and red box-plots, respectively, represent the distribution of the corrected and uncorrected MSE for the given $\ell$. The images in inset, from top to bottom, represent examples of a distorted intensity pattern, the corresponding updated phase mask at the SLM, and and the turbulence-corrected image for $\pm \ell$\,=\,1, 3, 5, 7, and 9, respectively.} 
\label{fig:Figure 3}
\end{figure}

Similarly, we run simulations with $\eta\,=\,275$ and a pre-trained CNN 
to measure the MSE with respect to number of iterations taken by the feedback loop. 
Here,  an output is given at every 10th iteration and its MSE measured, as shown in Fig. \ref{fig:Figure 2} \textbf{(b)}. We find the MSE of the network decreases with the number of iterations up to 700, then remains effectively constant. 
A minimum MSE of 0.116 is reached at the 740th iteration, with an initial maximum MSE of 1.033 (without the GDO).  Thus, we choose 700 iterations at an MSE of 0.119 and standard deviation of 0.062, as the optimized rank of iteration when accounting for both accuracy and computational efficiency. 

While the accuracy of the CNN often depends on various parameters, here we focus on optimizing the number of training images per class. To achieve this, a (randomly) selected set of 50, 100, 150, 200, 250, 300, 350, 400, 450, 500, and 550 images out of the total 600 images for each training class are separately used to train the CNN. The simulations are then performed with the optimized network parameters, and the accuracy is calculated with respect to number of images per training class, as shown in Fig. \ref{fig:Figure 2} \textbf{(c)}. The ``Zero'' number of training images corresponds to the MSE without the CNN in the feedback network as shown by the ``red-blue'' dot. A maximum MSE of 2.271 with a standard deviation of 0.343 is found at zero training images (without the CNN in the feedback network). We note the sharp decline of the MSE to 0.395 with a pre-trained CNN consisting of only 50 images per training class. This demonstrates the strength of a CNN in the optimization loop. The resulting image profiles are visibly much closer to the desired ``petal'' patterns.  The MSE is found to slowly decrease up to 200 training images, after which it begins to oscillate, which is a sign of over-fitting. We reach an MSE of 0.117 with standard deviation of 0.063 for 200 images per training class, which we then set as the optimized number of training images in the following simulations. 

With the feedback network optimized, we now turn to the turbulence-correction simulations.  The network parameters are set to: $\eta\,=\,275$, number of iterations = 700, and number of images per training class = 200.   Turbulent propagation is simulated, and then corrected by the feedback network, for superposition OAM modes ranging from $\pm \ell$\,=\, 0 (Gaussian) to $\pm 10$. Here, we train the CNN for each value of $
\pm \ell$ separately. Also, note that we simulate 100 turbulence phase screens with the same strength, 
but with different and random phase distributions for each value of $\pm \ell$. Examples of such phase screens are shown Fig. \ref{fig:Figure 3} \textbf{(c)}.  
This results in 100 different test intensity images at the receiver for each value of $\pm \ell$. 
We find that our network efficiently updates the phase mask for the SLM and modifies the intensity pattern for a given unknown random turbulence such that a more accurate mode is detected at the receiver of the communication channel (typically an order-of-magnitude reduction in the MSE). Examples of uncorrected and distorted (left),  the corresponding corrected phase masks at the SLM (center), and corrected intensity patterns at the receiver (right), are shown in Fig. \ref{fig:Figure 3} \textbf{(d, e, f, g)}. While there is an expected increasing trend of MSE with increase in the mode index of OAM, the feedback network continues to efficiently correct for the turbulent propagation, as shown in Fig. \ref{fig:Figure 3} \textbf{(h)}. The whiskers on the box-plots represent MSE values between 10th and 95th percentile. The mean and notch, respectively, represent the average and median of the MSE for 100 corrected test intensity images at the receiver. We obtain a minimum MSE of 0.100 and a maximum MSE of 0.532 for the Gaussian mode ($\pm\ell$\,=\,0) and superposition mode with $\pm\ell$\,=\,10, respectively, with corresponding standard deviations of 0.064 and 0.252, for the given turbulence strength. We reiterate here that the random turbulence phase maps used are not contained in the training set.
\begin{figure}[t!]
\centering\includegraphics[width=.92\linewidth]{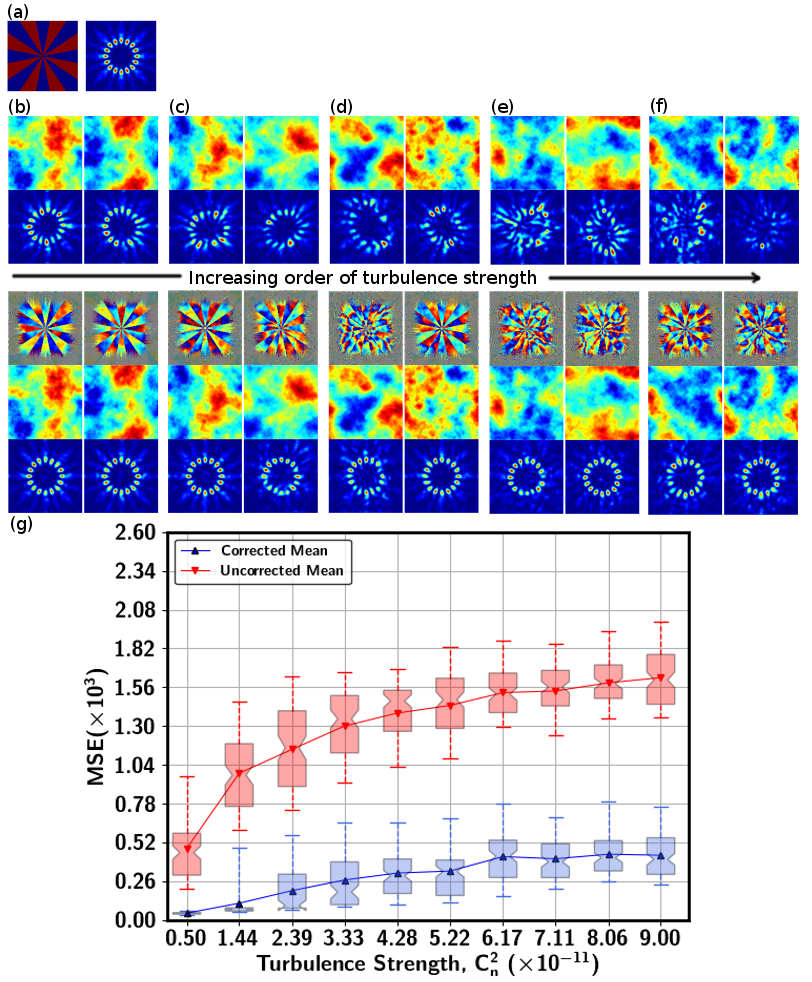}
\caption{\textbf{(a)} Ideal phase mask at the SLM (left) and target OAM image (without turbulence) at the receiver (right) for $\pm \ell\,=\, 8$. \textbf{(b, c, d, e, f)}: top two rows: turbulence phase screens (two screens per $C_n^2$) with varying strengths and the respective distorted intensity patterns at the receiver (without correction). bottom three rows: updated phase masks at the SLM, the same random phase screens, and the corrected intensity patterns at the receiver for the given strength of turbulence. \textbf{(g)} MSE versus turbulence strength for correcting the propagation of OAM superposition modes of $\ell\,=\,\pm 8$. The blue and red box-plots, respectively, represent the distribution of the corrected and uncorrected MSE for the given strength of turbulence. } 
\label{fig:Figure 4}
\end{figure}

Finally, the feedback network's ability to correct for turbulence of varying strength, $C_n^2$, is demonstrated. With the network settings optimized, the simulations are run and the corresponding MSE found, for images with OAM index $\ell$\,=\, $\pm 8$.
Here, we simulate 10 different test sets with turbulence strength uniformly chosen in a given range of $C_n^2$ (Table \ref{tab: table1}). 
Note that each test set has 100 random output intensity patterns from 100 random phase screens of the given strength. Examples (two screens per class) of test turbulence phase screens 
and the corresponding intensity outputs are shown in top two rows of Fig. \ref{fig:Figure 4} \textbf{(b, c, d, e, f)}. 
We find that our network successfully restores the lost spatial properties of the OAM mode at the receiver even in the presence of substantially strong turbulence, which are shown in the bottom three rows of Fig. \ref{fig:Figure 4} \textbf{(b, c, d, e, f)}. As expected, the network gives a increasing trend of the MSE (a decreasing trend of accuracy) with increasing turbulence strength as shown in \ref{fig:Figure 4} \textbf{(g)}.  Again, the feedback network successfully improves the MSE in a consistent manner. The whisker, mean, and notch, again, holds the same definition as described above.   
Here, we find a minimum MSE of 0.049 and a maximum MSE of 0.443  for turbulence strengths of $0.5\times10^{-11}$m$^{-2/3}$ and $8.06\,\times\,10^{-11}$ m$^{-2/3}$, respectively, with corresponding standard deviations of 0.008 and 0.163. As an example of the strength of the machine learning and GDO feedback network in turbulence correction for optical communications, the distorted images shown in \ref{fig:Figure 4} \textbf{(f)} (top) have spatial intensity profiles that are almost completely destroyed, whereas the corrected images shown below them are nearly identical to the desired intensity profile at the receiver.

In conclusion, we have demonstrated the power of convolutional neural networks in a feedback scheme to actively correct for the destructive effects of turbulent propagation on optical modes.  By modifying the transmitted mode profiles of the states to be sent through turbulence, received mode profiles are found to be nearly identical to the target, desired modes.  We hope that the present results will be directly applicable to increasing the robustness of both classical and quantum optical communications schemes.


\section*{Funding Information}
We acknowledge funding from the Louisiana State Board of Regents Research Competitiveness Subprogram under grant number 073A-15, as well as from Northrop Grumman -- NG NEXT.

\bibliography{Turbulence_paper}


\end{document}